\documentstyle[12pt]{article}

\input epsf

\begin{document}
\baselineskip=20.5pt

\def\vec#1{{\bf #1}}
\def\v#1{{\bf #1}}
\def\beqra{\begin{eqnarray}} \def\eeqra{\end{eqnarray}}
\def\beqast{\begin{eqnarray*}}
\def\eeqast{\end{eqnarray*}}
\def\beq{\begin{equation}}      \def\eeq{\end{equation}}
\def\be{\begin{enumerate}}   \def\ee{\end{enumerate}}

\def\fnote#1#2{\begingroup\def\thefootnote{#1}\footnote{
#2}
\addtocounter
{footnote}{-1}\endgroup}

\def\itp#1#2{\hfill{NSF-ITP-{#1}-{#2}}}

\def\gam{\gamma}
\def\Gam{\Gamma}
\def\la{\lambda}
\def\eps{\epsilon}
\def\La{\Lambda}
\def\si{\sigma}
\def\Si{\Sigma}
\def\al{\alpha}
\def\Tha{\Theta}
\def\tha{\theta}
\def\vphi{\varphi}
\def\del{\delta}
\def\Del{\Delta}
\def\ab{\alpha\beta}
\def\om{\omega}
\def\Om{\Omega}
\def\mn{\mu\nu}
\def\mun{^{\mu}{}_{\nu}}
\def\kap{\kappa}
\def\rsi{\rho\sigma}
\def\beal{\beta\alpha}
\def\ups{\upsilon}

\def\til{\tilde}
\def\rta{\rightarrow}
\def\eqv{\equiv}
\def\nab{\nabla}
\def\pa{\partial}
\def\sit{\tilde\sigma}
\def\ul{\underline}
\def\indt{\parindent2.5em}
\def\nd{\noindent}

\def\rsi{\rho\sigma}
\def\beal{\beta\alpha}

\def\caa{{\cal A}}
\def\cb{{\cal B}}
\def\cac{{\cal C}}
\def\cd{{\cal D}}
\def\ce{{\cal E}}
\def\cf{{\cal F}}
\def\cg{{\cal G}}
\def\cah{{\cal H}}
\def\ci{{\cal I}}
\def\cj{{\cal{J}}}
\def\ck{{\cal K}}
\def\cl{{\cal L}}
\def\cm{{\cal M}}
\def\cn{{\cal N}}
\def\cO{{\cal O}}
\def\cp{{\cal P}}
\def\car{{\cal R}}
\def\cs{{\cal S}}
\def\ct{{\cal{T}}}
\def\cu{{\ca{U}}}
\def\cv{{\cal{V}}}
\def\cw{{\cal{W}}}
\def\cx{{\cal{X}}}
\def\cy{{\cal{Y}}}
\def\cz{{\cal{Z}}}

\def\raisenot{\raise .5mm\hbox{/}}
\def\nota{\ \hbox{{$a$}\kern-.49em\hbox{/}}}
\def\notA{\hbox{{$A$}\kern-.54em\hbox{\raisenot}}}
\def\notb{\ \hbox{{$b$}\kern-.47em\hbox{/}}}
\def\notB{\ \hbox{{$B$}\kern-.60em\hbox{\raisenot}}}
\def\notc{\ \hbox{{$c$}\kern-.45em\hbox{/}}}
\def\notd{\ \hbox{{$d$}\kern-.53em\hbox{/}}}
\def\notbd{\ \hbox{{$D$}\kern-.61em\hbox{\raisenot}}} 
\def\note{\ \hbox{{$e$}\kern-.47em\hbox{/}}}
\def\notk{\ \hbox{{$k$}\kern-.51em\hbox{/}}}
\def\notp{\ \hbox{{$p$}\kern-.43em\hbox{/}}}
\def\notq{\ \hbox{{$q$}\kern-.47em\hbox{/}}}
\def\notW{\ \hbox{{$W$}\kern-.75em\hbox{\raisenot}}}
\def\notz{\ \hbox{{$Z$}\kern-.61em\hbox{\raisenot}}}
\def\notpa{\hbox{{$\partial$}\kern-.54em\hbox{\raisenot}}}

\def\fo{\hbox{{1}\kern-.25em\hbox{l}}}  
\def\rf#1{$^{#1}$}
\def\bx{\Box}
\def\tr{{\rm Tr}}
\def\rmtr{{\rm tr}}
\def\dag{\dagger}

\def\lag{\langle}
\def\rag{\rangle}
\def\bmid{\big|}

\def\vlap{\overrightarrow{\La p}} 
\def\lrta{\longrightarrow}
\def\lrar{\raisebox{.8ex}{$\longrightarrow$}}
\def\rlarw{\longleftarrow\!\!\!\!\!\!\!\!\!\!\!\lrar}

\def\llra{\relbar\joinrel\longrightarrow}     
\def\mapright#1{\smash{\mathop{\llra}\limits_{#1}}}
\def\mapup#1{\smash{\mathop{\llra}\limits^{#1}}}
\def\asymptotic{{_{\stackrel{\displaystyle\longrightarrow}
{x\rightarrow\pm\infty}}\,\, }} 
\def\asymptext{\raisebox{.6ex}{${_{\stackrel{\displaystyle\longrightarrow}
{x\rightarrow\pm\infty}}\,\, }$}} 

\def\7#1#2{\mathop{\null#2}\limits^{#1}}   
\def\5#1#2{\mathop{\null#2}\limits_{#1}}   
\def\too#1{\stackrel{#1}{\to}}
\def\tooo#1{\stackrel{#1}{\longleftarrow}}
\def\nout{{\rm in \atop out}}

\def\one{\raisebox{.5ex}{1}}
\def\BM#1{\mbox{\boldmath{$#1$}}}

\def\ltsim{\matrix{<\cr\noalign{\vskip-7pt}\sim\cr}}
\def\gtsim{\matrix{>\cr\noalign{\vskip-7pt}\sim\cr}}
\def\haf{\frac{1}{2}}


\def\place#1#2#3{\vbox to0pt{\kern-\parskip\kern-7pt
                             \kern-#2truein\hbox{\kern#1truein #3}
                             \vss}\nointerlineskip}

\def\illustration #1 by #2 (#3){\vbox to #2{\hrule width #1
height 0pt
depth
0pt
                                       \vfill\special{illustration #3}}}

\def\scaledillustration #1 by #2 (#3 scaled #4){{\dimen0=#1
\dimen1=#2
           \divide\dimen0 by 1000 \multiply\dimen0 by #4
            \divide\dimen1 by 1000 \multiply\dimen1 by #4
            \illustration \dimen0 by \dimen1 (#3 scaled #4)}}

\def\ON{{\cal O}(N)}
\def\UN{{\cal U}(N)}
\def\bdPh{\mbox{\boldmath{$\dot{\!\Phi}$}}}
\def\bPh{\mbox{\boldmath{$\Phi$}}}
\def\bPhs{\bPh^2}
\def\sef{S_{eff}[\sigma,\pi]}
\def\sigx{\sigma(x)}
\def\pix{\pi(x)}
\def\bph{\mbox{\boldmath{$\phi$}}}
\def\bphs{\bph^2}
\def\ex{\BM{x}}
\def\exs{\ex^2}
\def\xdot{\dot{\!\ex}}
\def\y{\BM{y}}
\def\ys{\y^2}
\def\ydot{\dot{\!\y}}
\def\pat{\pa_t}
\def\pax{\pa_x}

\renewcommand{\theequation}{\arabic{equation}}



\vspace*{.3in}
\begin{center}
 \large{\bf Topological Degeneracy of Quantum Hall Fluids}
\normalsize

\vspace{36pt}
X.G. Wen$^{a,b}$ \& A. Zee$^c$ \\
\vspace{20pt}
a) Department of Physics, \\
Massachusetts Institute of Technology,\\
Cambridge, MA 02139, USA \\
b) Physics Division, \\
National Center for Theoretical Sciences\\
P.O.Box 2-131, Hsinchu, Taiwan 300, ROC\\
c) {{Institute for Theoretical Physics,}\\
{University of California,\\ Santa Barbara, CA 93106, USA}\\}

\vskip 3mm
\begin{abstract}
We present a simple approach to calculate the degeneracy and the structure of
the ground states of non-abelian quantum Hall (QH) liquids on the torus. Our
approach can be applied to any
QH liquids (abelian or non-abelian) obtained from the parton construction. We
explain our approach by studying a series of examples of increasing complexity.
When the effective theory of a non-abelian QH liquid is a non-abelian
Chern-Simons (CS) theory, our approach reproduces the well known results
for the ground state degeneracy of the CS theory. However, our approach also
apply to  non-abelian QH liquids whose effective theories are not known and
which cannot be written as a non-abelian CS theory. We find that the ground
states
on a torus of
all  non-abelian QH liquids obtained from the parton construction
can be described by points on a lattice inside a
``folded unit cell." The folding is generated by reflection,
rotations, etc. Thus the ground state structures on the torus described
by the ``folded unit cells'' provide a way to (at least partially) classify
non-abelian QH liquids obtained from the parton construction.

\end{abstract}

\end{center}
\vskip 2mm
\vspace{.6cm}


\vfill
\pagebreak

\setcounter{page}{1}

\section{Introduction}

It has become increasingly clear that the Quantum Hall (QH)
liquids\cite{L, laugh}
are not merely disordered liquids: they
are quantum liquids with extremely rich and totally new
internal structures (or topological orders)\cite{top}.
Thus QH liquids represent a new
class of matter. Although we still do not have a complete theory of this new
kind of order, we do know that QH liquids can be divided into two
classes -- abelian\cite{GL, rev1, rev} and non-abelian\cite{nab,nabw}.  The
effective theory of
abelian QH liquids is known to be the $U(1)$ Chern-Simons (CS)
theory \cite{rev1, rev}, and because
of that we have a classification of all abelian QH liquids \cite{class} in
terms of the so-called $K$ matrix.

In contrast, the effective theories of many known non-abelian QH liquids
are unknown.
The problem is not because we
know too little about a non-abelian QH liquid so
that we cannot deduce the effective theory. In many cases, we know a lot
about the low energy properties of a QH liquid, and still do not know its
effective theory. This is
simply because the correct effective theory has not been named yet.
Giving a name is easy, but giving a proper name is hard.
Giving a proper name which carries meaningful information amounts to the
task of classifying non-abelian QH liquid, and so far we do not know how to
do this.

In this paper we concentrate on the physical properties of non-abelian states
on the torus, and use a simple approach to calculate the ground
state degeneracy of non-abelian states. We assume no prior knowledge of the
subject. Our approach is down to earth, and we proceed through specific
examples.

First we test our approach on abelian
states.
Then we study the non-abelian
state with wave function $(\chi_q(\{z_i\}))^2$ where $\chi_q$ is the wave
function with $q$ filled Landau levels. Using the $SU(2)_q$ CS effective
theory of the $(\chi_q({z_i}))^2$ state\cite{nabw},
we find that
the $(\chi_q({z_i}))^2$ state has $q+1$ degenerate ground states on a torus.
The $q+1$ degeneracy of the  $SU(2)_q$ CS theory has been calculated before,
using a powerful mathematical approach based on algebraic geometry, topological
theory, and Lie algebraic theory \cite{cs1, coset}. We hope that our discussion
will be more accessible to the non-mathematical reader.
Next, we study a slightly more complicated non-abelian
state with wave function $\chi_1(\{z_i\})(\chi_q(\{z_i\}))^2$.
We are able to set up a simple model which describes the $(q+1)(q+2)/2$
degenerate ground
states of the $\chi_1(\{z_i\})(\chi_q(\{z_i\}))^2$ QH liquid on the torus.
It is not clear, however, whether our model can be derived from a named
topological field theory.
Since the model is given by the $U(1)_{2q+4}\times SU(2)_q$
CS theory (which has $2(q+1)(q+2)$ ground states on the torus) with some 
additional projections, we call our
effective theory $(U(1)_{2q+4}\times SU(2)_q)/Z_2$ theory.
Then, we study the non-abelian
state with wave function $[\chi_q(\{z_i\})]^3$
(associated with $SU(3)$ CS theory)
to demonstrate more features of the ground states on torus.
In particular we show that, on torus, the non-abelian state 
$[\chi_q(\{z_i\})]^3$ can described by a $U(1)\times U(1)$ CS theory 
plus some projections.

Although we only discussed some simple examples, the approach used here can be
applied to more complicated non-abelian states. Through those studies we
see some general patterns. We hope those patterns will shed light on
how to classify topological orders in
non-abelian QH liquids. We outline such a classification in the concluding
section.

\section{Abelian FQH states}

It is well known by now that the fractional quantum Hall (FQH) fluid can be
represented
effectively by a Chern-Simons CS field theory \cite{rev1,rev, class}
\beq\label{cs}
S ~=~ \int ~d^2x~dt ~ {1 \over 4\pi} ~K_{IJ} ~\eps^{\mu \nu \la} ~ a_\mu^I
~\partial_\nu ~a_\la^J
\eeq
with an integer valued matrix $K$. The theory is topological,
with low energy properties characterized by degenerate ground states.
In a classic paper \cite{wen} on the
subject, Wen showed that the ground state degeneracy of (\ref{cs}) when
quantized on a manifold of genus $g$ is given by
\beq\label{gen}
D ~=~ (det ~K)^g
\eeq

In this paper, we study the simple case when the manifold is a torus
$(g=1)$. First let we assume the matrix $K$ is $1\times1$ and equal to an
integer $k$. In this case the CS theory (\ref{cs}) is the effective theory of
filling fraction $\nu=1/k$ Laughlin state, and $D$ is just the ground
state degeneracy of the Laughlin state. Let us call the CS theory with
$K=k$ the $U(1)_k$ CS theory.
The approach used below and in the next section for abelian and non-abelian
CS theories are not entirely
new (see Ref. \cite{cs1}). We present it here for the
purpose of introducing proper notations and concepts for later discussions.
We also present our introduction to the CS theory (abelian or non-abelian)
in a way which is easy for people not in the field
of field/string theory to understand.

Wen \cite{wen} determined the ground state degeneracy $D$ by adding to $S$
the non-topological Maxwell term
\beq\label{max}
\int ~d^2x ~dt ~\sqrt{g} ~g^{\mu \la} ~g^{\nu \sigma} ~{1 \over f^2}
~F_{\mu \nu}~F_{\la \sigma}
\eeq
with $f^2$ some coupling constant. Here $g^{\mu \nu}$ denote the metric of
the manifold.  On a torus, the ground state properties are determined by
constant (independent of $(x_1,x_2)$ but of course dependent on time) gauge
potentials
\begin{equation}
\label{xy}
 a_0(x_1,x_2,t)=0,\ \ \ a_1(x_1,x_2,t)=\frac{2\pi x(t)}{L_1},\ \ \
 a_2(x_1,x_2,t)=\frac{2\pi y(t)}{L_2}
\end{equation}
where $(L_1,L_2)$ are the size of the torus. The dynamics of the constant
gauge potentials is described by the Lagrangian obtained by inserting
(\ref{xy})
into (\ref{cs}) and (\ref{max}) (with $m$ determined by the Maxwell
coupling $1/f^2$)
\begin{equation}
\label{Lm}
L=\pi k ( y\dot{x} -x \dot{y}) +\frac12
m \left( \dot{x}^2 +  \dot{y}^2 \right)
\end{equation}
The ``large" gauge transformations $V ~=~e^{i 2\pi x_1/L_1}$ and $V ~=~
e^{i 2\pi x_2/L_2}$
transform a constant gauge potential to another constant gauge potential, and
induce (under the gauge transformation $a_\mu ~\rta~a_\mu
~-~iV^{-1}\partial_\mu V$) the following changes: $(x,y)\to (x+1,y)$ and
$(x,y)\to (x,y+1)$.
Since $(x,y)$, $(x+1,y)$, and $(x,y+1)$ are related by gauge transformation,
they represent equivalent point
\begin{equation}
\label{equiv}
 (x,y) \sim (x+1,y) \sim (x,y+1)
\end{equation}
Thus Lagrangian in (\ref{Lm}) describes a mass $m$ particle on a torus
with a uniform
``magnetic field.'' The total number of flux quanta is $k$, which leads to a
$k$-fold degenerate ground state.

The defining characteristic of a topological theory such as
(\ref{cs}) is that it does not depend on the metric.
We should be able to determine the ground state of a topological field
theory without
having to add a regulating term which breaks the topological character of
the theory.  In the following, we will determine the ground state
degeneracy without adding the regulating term.

Given (taking $m=0$ in (\ref{Lm}))
\beq\label{one}
S ~=~ \int ~dt~L ~=~ \int ~ dt ~ 2 \pi k~x~\dot{y}
\eeq
we have
\beq\label{can}
{\delta L \over \delta \dot{y}} ~=~ 2 \pi ~kx
\eeq
and so
\beq\label{one*}
\left[x,y \right] ~=~ {i \over 2 \pi k}
\eeq
If we regard $y$ as the position
variable, the conjugate momentum is given by $p ~=~ 2 \pi k x$. The
Hamiltonian $H$
vanishes,
\beq\label{ham}
H ~=~ p~\dot{y} ~-~ L ~=~ 0
\eeq
a hallmark of a topological theory. Thus the Schr\"odinger
equation just reads
\beq
0 \cdot \psi~=~E\psi
\eeq
How then do we determine the wave functions $\psi(y)$?

Naively, any function $\psi(y)$ will have zero energy and would qualify as
a ground state wave
function. But such a wave function would in general not satisfy the
equivalence condition
(\ref{equiv})! The
allowed wave functions are determined by the requirement
that the particle lives on a torus with coordinates $(x,y)$ such that $x
\sim x+1$ and $y \sim y+1$. In other words,
$y$ and $y+1$ really represent the same point and hence we must require
$\psi (y) ~=~ \psi(y+1)$. This periodicity condition implies that
\beq\label{two}
\psi(y) ~=~ \sum_{n~=~-\infty}^{\infty}c_n~e^{i 2 \pi n y}
\eeq
with $n$ an integer. To impose the  periodicity condition in the $x$ direction
$x \sim x+1$, we need to make a Fourier transformation to obtain the wave
function
in the ``momentum'' space. Canonical conjugation gives $p ~=~ i {\partial \over
\partial y}$
and so
\beq
\tilde\psi(p) ~=~ \sum c_n ~ \delta(p ~-~ 2 \pi ~n)
\eeq
It is useful (if only to get rid of the $(2 \pi)$'s, but more importantly
to emphasize
the equal status of $x$ and $y$) to recall $p~=~2 \pi ~k~x$ and hence to
define a wave function in the $x$ coordinate
\beq\label{john}
\phi(x) ~=~ \sum c_n ~ \delta (kx ~-~ n)
\eeq

The condition $x \sim x+1$  now implies that
\beq\label{nk}
c_n ~=~ c_{n+k}
\eeq
We have thus reached the conclusion that the ground state degeneracy $D$ is
$k$, since there are $k$ independent $c_n$'s, namely $c_1, c_2, \ldots,
c_k$. We have thus shown how to determine the ground state degeneracy
without breaking the topological character of the theory.

It is useful to
define the periodic delta function
\beq\label{del}
\delta^P (y) ~\equiv~ \sum_{l=-{\infty}}^{\infty}~e^{i2 \pi ly}
\eeq
equal to $1$ (up to some irrelevant overall infinite constant) if $y$ is an
integer, and $0$ otherwise.
It is also convenient to write, for any integer $n$,
\beq
 n ~=~lk ~+~ m
 \eeq
(with $l$ an integer) and define $[n]_k ~\equiv~m$ as the reduced part of $n$.
We will suppress the index $k$ on $[n]_k$ if there is no ambiguity. With
these definitions we can re-write (\ref{two}) as
\beqra\label{four}
\psi(y)~ &=~ &\left(\sum_{m=1}^k ~c_m ~ e^{i 2\pi my}
\right)~\left(\sum_{l=-{\infty}}^{\infty}~e^{i2 \pi lky}\right) \nonumber \\
&=~ & g(y) ~ \delta^P(ky)
\eeqra
Again writing
$n ~=~ lk ~+~ m$, we obtain from (\ref{john}) that $[kx] ~=~m$. Since
$c_n ~=~ c_m$ by periodicity, we have
\beq\label{five}
\phi(x) ~=~ c_{[kx]} ~ \delta^P(kx)
\eeq
This shows that $\phi(x)$ and $\psi(y)$ are indeed on the same footing,
and that $\psi(y)$ is, in the sense described here, just the Fourier transform
of $c_m$. We note
that $\phi(x)$ and $\psi(y)$ are proportional to $\delta^P (kx)$ and
$\delta^P (ky)$ respectively.
Thus the position coordinates $x$ and $y$ are both quantized
to be ${1 \over k}, {2 \over k}, \ldots, {k-1 \over k}, 1$.
Of course, $x$ and $y$ cannot be simultaneously diagonalized. More
precisely, the wave function $\psi(x)$ is non-zero only when $x$ take on
certain discrete values, (and similarly with $x$ replaced by $y$.)

We mention here for later use that it is now of course straightforward to
generalize
the case given in (\ref{cs}). The effective point particle quantum mechanics is
described by the Lagrangian
\beq\label{st1}
L ~=~ 2\pi~K_{IJ}~x_I~\dot{y}_J
\eeq
\noindent (repeated indices in $I,J$ summed) and the commutation relation
\beq\label{st2}
\left[x_I,~y_J\right] ~=~ {i \over 2 \pi}~(K^{-1})_{IJ}
\eeq

\noindent The wave function in the $y$ coordinates is given by

\beq\label{ycoord}
\psi(\vec{y}) ~=~ \sum_{\vec{n}} ~c_{\vec{n}} ~e^{i 2 \pi ~\vec{n} \cdot \vec{y}}
\eeq

\noindent where $n_I$, the components of $\vec{n}$, are integers. The
corresponding
wave function in the $x$-coordinates is

\beq\label{xcoord}
\psi(\vec{x}) ~=~ \sum_{\vec{n}} ~c_{\vec{n}} ~\delta\left(x_I ~-~ (K^{-1})_{IJ}
~n_J\right)
\eeq

\section{$SU(2)$ non-abelian FQH states}

Next, let us calculate the ground
states degeneracy of a simple class of non-abelian QH states.
It was pointed out \cite{nabw} that the QH liquid described by wave function
$ \left(\chi_q(z_1,...,z_N) \right)^2$ (where $\chi_q$ is the fermion wave
function with $q$ filled Landau levels) is a non-abelian QH state, whose
effective theory is the $SU(2)_q$ CS theory ({\it i.e.} the $SU(2)$ level-$q$
CS theory). Let us first recall how the non-abelian states \cite{nab} can be
constructed, using the rather physical parton construction \cite{nabw}. For
the
sake of
pedagogical clarity, we focus on a specific (but unphysical) example.

We imagine that at short distances the electron
can be
cut into two constituents (``partons"),
each of charge $e_0 ~=~ e/2$.
The long distance physics of the
resulting Hall fluid should
be independent of the details of the short
distance dynamics. This can
be explained very simply in terms of wave
functions. Denote the
coordinates of the partons by $\{z_i^{\alpha}\}
~, ~\alpha ~=~ 1,2$ and $i=1,2, \dots, N~$, with
$N$ electrons in the
system. Let each species of partons fill $q$
Landau
levels, and denote the corresponding wave
function by $\chi_q$. The
wave function of the entire fluid is then given
by
\beq\label{fluid} \Psi ~\sim ~\chi_q
~\left(z_1^{(1)}, \dots,
z_N^{(1)}\right) ~ \chi_q~\left(z_1^{(2)},
\dots, z_N^{(2)}\right)
\eeq
We now have to tie the ``partons" together to
form the electrons. This
is done by setting $z_j^{(1)}~=~z_j^{(2)}~=~
 z_j$ in the
wave function $\Psi$. For instance, for $q=1$,
we have $\chi_1 ~=~
\prod\limits_{i>j} (z_i~-~ z_j)$, and so we
obtain $\Psi ~\sim~
\prod\limits_{i>j} (z_i ~-~ z_j)^2$, which is
nothing but the $\nu ~=~ {1\over
2}$ Laughlin state (for bosonic electrons), as Laughlin \cite{laugh} taught
us.

In field theoretic language, before we bind the
``partons" together into
electrons, we have the Lagrangian
\beqra
\cl &= &i \psi^{\dag}_1 \pa_t \psi_1 ~+~ {1
\over 2m}
\psi^{\dag}_1(\pa_i-i e_0 A_i)^2 \psi_1
\nonumber \\
&&+ i \psi^{\dag}_2 \pa_t \psi_2 ~+~ {1 \over
2m} \psi^{\dag}_2(\pa_i-i e_0
A_i)^2 \psi_2
\eeqra
with $\psi_1$, $\psi_2$
corresponding to the two parton fields. We glue
the
two ``partons" together by coupling them to an
$SU(2)$ gauge potentials
$a_\mu$:
\beq\label{55a}
\cl~=~ i \psi^{\dag} (\pa_t -i a_0) \psi + {1
\over 2m}
\psi^{\dag}(\pa_i-i e_0 A_i-i a_i)^2 \psi
\eeq
where $\psi=\pmatrix{\psi_1\cr \psi_2\cr
}$ and $a_\mu$ are
hermitian traceless 2 by 2 matrices. Now we
can integrate out $\psi_{1,2}$
(see for example Ref. \cite{rev}) and obtain the
effective theory:
\beq\label{SU2CS}
\cl = \frac{2 q e_0^2}{4\pi}~\eps^{\mu\nu\la} A_\mu\pa_\nu
A_\la +
\frac{q}{4\pi}{\rm Tr}\eps^{\mu\nu\la}
(a_\mu \pa_{\nu} a_\la + \frac{i}{3}
a_\mu a_\nu a_\la)
\eeq
All we need from Ref. \cite{rev} is the result that given

\beq\label{xx}
\cl ~=~ i \psi^{\dag} \pa_t \psi ~+~ {1 \over
2m} \psi^{\dag} (\pa_i-i e_0
A_i)^2 \psi
\eeq
and with $\psi$ filling $q$ Landau levels we obtain the effective Lagrangian

\beq\label{xx2}
\cl ~=~ {qe^2_0 \over 4 \pi}~\eps^{\mu \nu \lambda}~A_\mu \pa_\nu A_\lambda
\eeq
upon integrating out $\psi$. There is also an analogous formula for the
non-abelian sector.

The first term tells us that the this state has a
filling fraction $\nu ~=~ 2q e_0^2
/e^2~=~q/2$. The second term
describes a level $q$ $SU(2)$ (denoted by
$SU(2)_q$) Chern-Simons CS
effective theory, which determines the ground
state properties of the QH liquid on compact
spaces.

Starting from the $SU(2)_q$ CS theory (\ref{SU2CS}),
we choose the $a_0=0$ gauge to calculate the ground states on a torus.
In this gauge we need to enforce the constraint of zero field strength:
$f_{ij}=0$. Introduce Wilson loop operators
$U_c \equiv P[e^{i\oint_c dx_\mu a_\mu}] \in SU(2)$
(where $P[...]$ is a path ordered
product). For a contractable path $c$, we have trivially $U_c=1$ due to the
constraint. On a torus, all
the gauge invariant quantities are contained in the two Wilson loop operators
for the two non-contractable loops $c_{1,2}$ in the $x_1$ and $x_2$ directions:
$U_1= P[e^{i\oint_{c_1} dx_\mu a_\mu}]$ and
$U_2=P[e^{i\oint_{c_2} dx_\mu a_\mu}]$.
Since $U_1 U_2 U_1^\dagger U_2^\dagger$ is a Wilson loop operator for a
contractable loop, we have $U_1 U_2 U_1^\dagger U_2^\dagger=1$, and $U_1$
commutes with $U_2$. Making a global $SU(2)$ gauge transformation, we can make
$U_{1,2}$ have the form
\begin{equation}
 U_1=e^{i2\pi x \tau_3},\ \ \ \ U_2=e^{i2\pi y \tau_3}
\end{equation}
This corresponds to
spatially constant gauge potentials: writing $a_i=a^l_i \tau_l$ with
$\tau_l$ the usual Pauli matrices we have
\begin{equation}
\label{a123}
 a^{1,2}_i =0,\ \ \ a^3_1(x_1,x_2,t)=2\pi x(t)/L_1,\ \ \
 a^3_2(x_1,x_2,t)=2\pi y(t)/L_2
\end{equation}
We see that the $SU(2)$ CS theory has at low energies non-trivial physical
degrees of
freedom described by $(x,y)$ (just as in (\ref{Lm})). Classically different
values of
$(x,y)$ correspond to different degenerate physical states, and there are
infinite number of degenerate ground states. However, in quantum theory,
$x$ and $y$ do
not commute and the uncertainty relation leads to only a finite number of
degenerate ground states. To obtain the commutator between $x$ and $y$, we
insert (\ref{a123}) into  (\ref{SU2CS})
to obtain the effective theory

\begin{equation}
\label{SSU2}
 S=\int dt \  2\pi q (x \dot y - y \dot x).
\end{equation}
This is identical to the effective theory for the $U(1)_k$
theory (\ref{one})
with the identification $k=2q$. The resulting commutator is
\begin{equation}
 [x,y]={i\over 2\pi k}
\end{equation}
Again, under a large gauge transformation, we have
\begin{equation}
\label{eSU2}
 x\sim x+1,\ \ \ y\sim y+1
\end{equation}

Thus, naively one might think that the $SU(2)_q$ theory is
described by
 (\ref{SSU2}) and (\ref{eSU2}) which is nothing but a $U(1)_k$ theory with
$k=2q$. But
that would be wrong. In fact, the $SU(2)_q$ theory is not equivalent to the
$U(1)_{2q}$ theory, because the $SU(2)$ CS theory contains an additional global
$SU(2)$ gauge
transformation that changes $a^3$ to $-a^3$.  This gauge
transformation imposes an additional equivalence condition
\begin{equation}
\label{e2SU2}
 (x,y)\sim (-x, -y)
\end{equation}
Eqs. (\ref{SSU2}), (\ref{eSU2}), and (\ref{e2SU2}) form a complete description
of the $SU(2)_q$ theory on the torus.

{}From the above discussion, we see that the $SU(2)_q$ states  can be obtained
from the $U(1)_{k=2q} $ states. The $k$ states in the $U(1)_k$ theory
is given by
\beq
\psi(y) ~=~ \sum c_n~e^{i 2 \pi n y}
\eeq
or
\beq
\phi(x) ~=~ \sum c_n ~ \delta (kx ~-~ n)
\eeq
with $c_n=c_{n+k}$. The additional condition (\ref{e2SU2}) implies that
only those states that satisfy $\psi(y) =\psi(-y)$ and $\phi(x)=\phi(-x)$
can belong to the $SU(2)_q$ theory. This requires $c_n=c_{-n}$. Thus the
$SU(2)_q$ CS theory, as well as the non-abelian FQH state described by
$(\chi_q)^2$, has $q+1$ degenerate ground states, corresponding to the
$q+1$
independent coefficients $c_0$, $c_1$, ... , $c_q$.

It turns out that $SU(2)_1$ CS theory represents a special case. We note
that when $q=1$, the requirement $c_n=c_{-n}$ does not remove any
states. This agrees with the known result that
the $SU(2)_1$ CS theory is equivalent to the $U(1)_2$ CS theory.

\section{$U(1)\times SU(2)$ non-abelian FQH states}

In this section we are going to discuss
a non-abelian state which is closely related to the one discussed above, but
which is physical. The electron is a fermion and we would like to split it
into
three
fermionic partons. Thus we first split an electron into three
different partons of electric charge $e_0= q/(q+2) $, $e_1=e_2= 1/(q+2)$ (so
that $e_0 + e_1 + e_2 ~=~ 1$),
and write the above wave function as
$\chi_1(z_1^{(0)},..,z_N^{(0)})
\chi_q(z_1^{(1)},...,z_N^{(1)}) \chi_q(z_1^{(2)},...,z_N^{(2)})$. This
non-abelian state has a wave function
$\chi_1\!(z_1,..,z_N)\!\! \left(\chi_q(z_1,...,z_N)\! \right)^2$.
The effective theory for the partons is given by
\begin{eqnarray}
 \cl &= &i \psi^\dag_0 \pa_t \psi_0
 +\frac{1}{2m} \psi^\dag_0(\pa_i-i e_0 A_i)^2 \psi_0 \nonumber\\
 &&+ i \psi^\dag_1 \pa_t \psi_1
 +\frac{1}{2m} \psi^\dag_1(\pa_i-i e_1 A_i)^2 \psi_1 \nonumber\\
 &&+ i \psi^\dag_2 \pa_t \psi_1
 +\frac{1}{2m} \psi^\dag_2(\pa_i-i e_2 A_i)^2 \psi_2
\end{eqnarray}
The above effective theory describes three independent QH fluids of filling
fraction $\nu_0=1$, $\nu_1=\nu_2=q$. Now we include a $U(1)$ and an
$SU(2)$ gauge field, $b_\mu$ and $a_\mu$, to recombine partons together
to form an electron:
\begin{eqnarray}
 \cl&=&
 i \psi^\dag_0 (\pa_t- 2i b_0) \psi_0
  +\frac{1}{2m} \psi^\dag_0(\pa_i-i e_0 A_i- 2i b_i)^2 \psi_0 \nonumber\\
 &&+ i \psi^\dag (\pa_t -i a_0 + i b_0 )\psi
 +\frac{1}{2m} \psi^\dag(\pa_i-i e_1 A_i-i a_i + ib_i )^2 \psi \nonumber \\
\end{eqnarray}
where $\psi=\pmatrix{\psi_1\cr \psi_2\cr}$ and $a_\mu$ are 2 by 2
matrices. Now we can integrate out $\psi_{0,1,2}$ (see (\ref{xx2}) or Ref.
\cite{rev}) and obtain the effective
theory for the $ \chi_1\chi_q^2$ state:
\begin{eqnarray}
\cl &=& \frac{e_0^2+2q e_1^2}{4\pi} A_\mu\pa_\nu A_\la \eps^{\mu\nu\la}+
 \frac{2^2+2q}{4\pi} b_\mu\pa_\nu b_\la \eps^{\mu\nu\la}+
 \frac{q}{4\pi}{\rm Tr}\eps^{\mu\nu\la} (a_\mu \pa_{\nu} a_\la +
\frac{i}{3} a_\mu a_\nu a_\la) \nonumber\\
&&- \frac{4e_0+4q e_1}{4\pi} A_\mu\pa_\nu b_\la \eps^{\mu\nu\la}
 \label{U1SU2}
\end{eqnarray}
The first term tells us that the $ \chi_1\chi_q^2$
state has a filling fraction $\nu=e_0^2+q e_1^2+q e_1^2=q/(q+2)$. The next two
terms describe a $U(1)_{2q+4} \times SU(2)_q$ CS effective theory, which
determines the ground state properties of the QH liquid on compact spaces.

According to the results we have thus far, the $U(1)_{2q+4} \times SU(2)_q$ CS
theory has $(2q+4)(q+1)$ degenerate
ground states on the torus.  Thus one may naively expect that the
$ \chi_1\chi_q^2$ state also has $(2q+4)(q+1)$ degenerate
ground states on the torus. However this results cannot be right, since
when $q=1$ the $ \chi_1\chi_q^2$ state is nothing but the $\nu=1/3$
Laughlin state and should have 3 degenerate
ground states instead of 12 as implied by $(2q+4)(q+1)$.
Therefore despite the above ``derivation'', the
$U(1)_{2q+4} \times SU(2)_q$ CS theory cannot be the correct effective theory
for the $ \chi_1\chi_q^2$ state. As we will see later, however,
the correct effective
theory can be obtained from the $U(1)_{2q+4} \times SU(2)_q$ CS theory.

Recall that the $U(1)_{2q+4} \times SU(2)_q$ CS theory on a torus can be
described by four degrees of freedoms $(x,y)$ and $(x',y')$. The corresponding
gauge fields are given by
\begin{equation}
 b_1(x_1,y_1, t)=2\pi \frac{x(t)}{L_1},\ \ \ \
 b_2(x_1,y_1, t)=2\pi \frac{y(t)}{L_2}
\end{equation}
and
\begin{equation}
 a_1(x_1,y_1, t)=2\pi \frac{x'(t)}{L_1}\pmatrix{1&0\cr 0&-1\cr},\ \ \ \
 a_2(x_1,y_1, t)=2\pi \frac{y'(t)}{L_2}\pmatrix{1&0\cr 0&-1\cr}
\end{equation}
This reduces the effective theory (\ref{U1SU2}) to
\begin{equation}
\label{SU1SU2}
L~=~\pi (2q+4)  (x \dot y - y \dot x)+2\pi q (x' \dot y' - y' \dot x').
\end{equation}
As noted before, large gauge transformations give us some equivalence
conditions.
For example the $U(1)$ large gauge transformation
$\exp \left(i \frac{2\pi x_1}{L_1} \pmatrix{2&&\cr &-1&\cr &&-1\cr} \right)$
that acts on $\pmatrix{\psi_0\cr \psi_1\cr \psi_2\cr}$ shifts $x$ to $x+1$.
This kind of large gauge transformations leads to the equivalence condition
\begin{equation}
\label{e1U1SU2}
x\sim x+1, \ \ \ \ y\sim y+1
\end{equation}
Similarly, the $SU(2)$ large gauge transformations (such as
$\exp \left(i \frac{2\pi x_1}{L_1} \pmatrix{0&&\cr &1&\cr &&-1\cr} \right)$)
leads to
\begin{equation}
\label{e2U1SU2}
 x'\sim x'+1,\ \ \ \ y'\sim y'+1
\end{equation}
The $SU(2)$ CS theory also has an additional reflection equivalence condition
\begin{equation}
\label{e3U1SU2}
 (x',y')\sim  (-x',-y')
\end{equation}
Equations (\ref{SU1SU2}), (\ref{e1U1SU2}),
(\ref{e2U1SU2}), and (\ref{e3U1SU2})
describe the
$U(1)_{2q+4} \times SU(2)_q$ CS theory on the torus and has $(2q+4)(q+1)$
degenerate ground states.

However, for our theory we have an additional
large gauge transformation which mixes the $U(1)$ and the center of $SU(2)$.
The large gauge transformations are given by
$\exp \left(i \frac{2\pi x_1}{L_1} \pmatrix{1&&\cr &-1&\cr &&0\cr} \right)$
and
$\exp \left(i \frac{2\pi x_2}{L_2} \pmatrix{1&&\cr &-1&\cr &&0\cr} \right)$
and gives rise to the following  equivalence conditions
\begin{equation}
\label{e4U1SU2}
 (x,x')\sim (x+\frac12, x'-\frac12),\ \ \ \
 (y,y')\sim (y+\frac12, y'-\frac12)
\end{equation}
since $\pmatrix{1&0\cr 0&0\cr} ~=~ {1 \over 2}~ \pmatrix{1&0\cr 0&1\cr} ~+~
{1 \over 2}~ \pmatrix{1&0\cr 0&-1\cr}$.

We believe that equations (\ref{SU1SU2}), (\ref{e1U1SU2}),
(\ref{e2U1SU2}), (\ref{e3U1SU2}), and (\ref{e4U1SU2})
describe the the correct effective theory for the $\chi_1 \chi_q^2$
state on torus. Because the correct effective theory is obtained from the
$U(1)_{2q+4}\times SU(2)_q$ CS theory by applying the additional equivalence
condition (\ref{e4U1SU2}), we will call it the $(U(1)_{2q+4}\times
SU(2)_q)/Z_2$
theory.
The edge excitations of the $\chi_1 \chi_q^2$ state is discussed in
Ref. \cite{nabw}, which is described by the $U(1)\times SU(2)_q$ KM algebra
matching very well with the bulk effective theory.

To obtain the ground state degeneracy of the $\chi_1 \chi_q^2$
state, we start with the $U(1)_{2q+4}\times U(1)_{2q}$ theory. The states
are given by
\begin{equation}
 \phi(x,x')=c_{[(2q+4)x], [2qx']} \del^P( (2q+4)x) \del^P(2qx')
\end{equation}
where the coefficient $c_{n,n'}$ (with $n,n'=$integer) satisfy
\begin{equation}
\label{ce1}
 c_{n,n'}=c_{n+2q+4,n'}=c_{n,n'+2q}
\end{equation}
as a consequence of the equivalence conditions (\ref{e1U1SU2}),
and (\ref{e2U1SU2}).
The condition $(x,x')\sim (x+\frac12, x'-\frac12)$ in
(\ref{e4U1SU2}) is satisfied by requiring
\begin{equation}
\label{ce2}
 c_{n,n'}=c_{n+q+2,n'-q}
\end{equation}
and $(y,y')\sim (y+\frac12, y'-\frac12)$ in
(\ref{e4U1SU2}) is satisfied by requiring
\begin{equation}
\label{ce3}
c_{n,n'}=0 \ \ \ \hbox{if  } n+n'=odd
\end{equation}
This can be seen from the relation
\begin{equation}
\psi(y,y') ~=~ \sum_{n, n'}c_{n,n'} ~e^{i 2 \pi (n y+n'y')}
\end{equation}
The reflection condition (\ref{e3U1SU2}) gives us
\begin{equation}
\label{ce4}
 c_{n,n'}=c_{n,-n'}
\end{equation}
In Fig. 1, the circles represent $(2q+4)(2q)/4$ independent $c_{n,n'}$
after imposing (\ref{ce1}), (\ref{ce2}) and (\ref{ce3}). The filled circles
represent
$\frac{(2q+4)(2q)}{4\times 2}+\frac{2q+4}{2\times 2}=\frac{(q+1)(q+2)}{2}$
independent $c_{n,n'}$ after imposing the additional reflection condition
(\ref{ce4}). Therefore the $\chi_1\chi_q^2$ state has

\beq\label{oneq}
D ~=~ (q+1)(q+2)/2
\eeq
degenerate ground states on torus.

\begin{figure}[htbp]
\epsfxsize=3in
\begin{center}
\leavevmode
\epsfbox{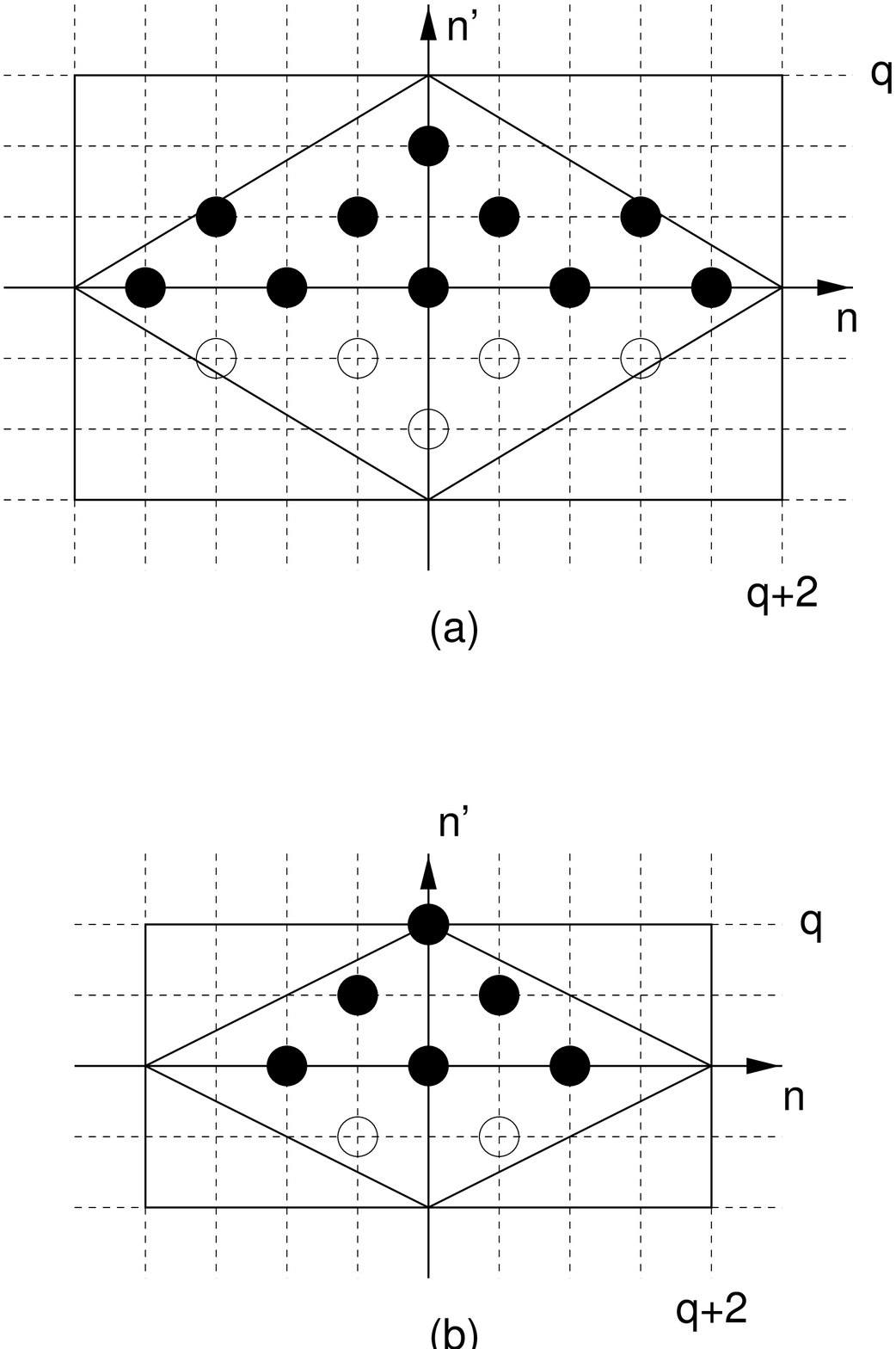}
\end{center}
\caption{The ground states of $\chi_1\chi_q^2$ non-abelian state for (a) $q=3$
and (b) $q=2$.}
\label{fig1}
\end{figure}

Note that when $q=1$ the $\chi_1\chi_q^2$
state is just the $\nu=1/3$ Laughlin state and $\frac{(q+1)(q+2)}{2}=3$
is the expected ground state degeneracy. When $q=2$ the $\nu=1/2$
$\chi_1\chi_2^2$ state has six degenerate ground states.

 By an elegant argument, Tao and Wu \cite{taowu}
 threaded a solenoid through the
 quantum Hall geometry proposed by Laughlin and showed that as the flux
 of the solenoid increases by quantized units, if the system goes through
 $p$ degenerate states before returning to its original state, an integer
 number $n$ of electrons are transported. Thus, the Hall conductance or
 filling fraction is given by $\nu=n/p$ while the degeneracy $D=m p$ is an
 integer $m$ multiple of $p$. That is
 the ground state degeneracy $D$ is an integer
 multiple of the denominator of $\nu$. 
We note that, for our results,
the denominator of the filling fraction $\frac{q}{q+2}$ is
always a factor of the degeneracy $\frac{(q+1)(q+2)}{2}$. This is consistent
with the above result.

\section{$SU(3)$ non-abelian FQH states}

Now let us calculate the ground
states degeneracy of $SU(3)$ non-abelian QH states, whose wave functions
are given by
$ \left(\chi_q(z_1,...,z_N) \right)^3$ and which have a filling fraction $q/3$.
In the parton construction, we cut the electron into three pieces, each of
charge $e_0 ~=~ e/3$. We put the partons into Landau levels. Finally, we glue
the partons together into electrons.

At the gluing stage, we have a choice. We can do the gluing either with an
$SU(3)$ gauge field, thus obtaining a non-abelian CS theory, or with two $U(1)$
gauge fields $a_\mu$ and $b_\mu$, thus obtaining an abelian
$\left(U(1)\right)^2$ CS theory. We will discuss the second option first as it
is conceptually somewhat simpler.

As before, we have

\beqra\label{glu}
\cl &= &i \psi^{\dag}_1 \left(\pa_t -
i(a_0+b_0)\right) \psi_1 ~+~ {1 \over 2m}
\psi^{\dag}_1(\pa_i-i e_0 A_i-i(a_i + b_i))^2
\psi_1 \nonumber \\
&&+ i \psi^{\dag}_2 \left(\pa_t -i(-
a_0+b_0)\right) \psi_2 ~+~ {1 \over 2m}
\psi^{\dag}_2(\pa_i-i e_0 A_i-i(-a_i + b_i))^2
\psi_2  \nonumber \\
&&+i \psi^{\dag}_3 \left(\pa_t -i(-
2b_0)\right) \psi_3 ~+~ {1 \over 2m}
\psi^{\dag}_3(\pa_i-i e_0 A_i-i(-2b_i))^2
\psi_3
\eeqra
(Note that we need two gauge fields. Suppose we
introduce only $b_\mu$. Then we
would have the bound states $\psi_1 \psi_1
\psi_3$, $\psi_2 \psi_2 \psi_3$, as
well as $\psi_1 \psi_2 \psi_3$. There would
be three different kinds of
electrons.)

Filling $q$ Landau levels with $\psi_1$,
$\psi_2$, $\psi_3$, and integrating out
the $\psi$ fields, we obtain
\beqra\label{susan}
\cl &=&~ {3qe^2_0 \over 4\pi} ~ \eps^{\mu \nu
\la} ~A_\mu \pa_\nu A_\la \nonumber \\
&&+~ {q \over 4 \pi}\eps^{\mu \nu \la} (2 a_\mu
\pa_{\nu} a_\la ~+~ 6 b_\mu
\pa_\nu b_\la)
\eeqra
Introducing Wilson loops as before, we
write

\beq\label{jill1}
a_1 ~=~ {2\pi \over L_1} x(t), ~ a_2 ~=~ {2\pi
\over L_2} y(t)
\eeq
and
\beq\label{jill2}
b_1 ~=~ {2\pi \over L_1} x'(t), ~ b_2 ~=~ {2\pi
\over L_2} y'(t)
\eeq
We insert (\ref{jill1}) and
(\ref{jill2}) into (\ref{glu}) to obtain the
effective action for the low energy degrees of
freedom:
\begin{equation} \label{sSU2} S=\int dt \  2\pi
q (x \dot y - y \dot x)
~+~ 6\pi q~(x'\dot{y'} ~-~ y' \dot{x'})
\end{equation}
Performing a $U(1)_a$ transformation of the
form $U_a ~=~ e^{i 2 \pi x_1 /
L_1}$, we conclude that $x ~\sim x+1$ with
$y,x', y'$ unchanged. Similarly,
performing a $U_b$ transformation, we
conclude that $x' ~\sim x'+1$ with $x,y,
y'$ unchanged. Furthermore, we can also
perform the corresponding transformations
along the $x_2$ direction and change $y$ and
$y'$ respectively. We can thus
interpret (\ref{sSU2}) as describing the motion
of two particles on a torus of size (1,1).

However, there is an additional slightly subtle
point: we can perform a gauge
transformation using the diagonal subgroup of
$U(1)_a ~\times~ U(1)_b$. More
precisely, we transform $\psi_1 \rta e^{i 2 \pi
x_1/ L_1} \psi_1$ and $\psi_3
\rta e^{-i 2 \pi x_1/ L_1} \psi_3$, leaving
$\psi_2$ unchanged. This implies
\beq\label{s}
S ~:~ (x,x') ~\sim~ \left(x ~+~ {1 \over 2}, ~ x'
~+~ {1 \over 2}\right)
\eeq
with $y,y'$ left invariant. Similarly, performing
the corresponding transformation
along the $x_2$ direction we have $(y,y')
~\sim~ \left(y ~+~ {1 \over 2}, ~ y' ~+~
{1 \over 2} \right)$. In other words, various
points in the phase space of the two
quantum particles have to be identified.

We can now write down the wave function in the
$y$ basis as
\beq\label{yb}
\psi(y,y') ~=~ \sum\limits_{n,n'} ~ c_{n,n'} ~
e^{i 2 \pi (n y~+~n' y')}
\eeq
and correspondingly in the $x$ basis as
\beq\label{xb}
\phi(x,x') ~=~ \sum\limits_{n,n'} ~ c_{n,n'} ~
\delta(2 q x ~-~n) ~\delta(6 qx'
~-~n')
\eeq
The torus boundary condition implies that $n$
and $n'$ are integers and
\beq\label{cst}
c_{nn'} ~=~ c_{n+2q,n'} ~=~ c_{n,n'+6q}
\eeq
Furthermore, the equivalence relation (\ref{s})
implies that
\beq\label{sh}
c_{nn'} ~=~ c_{n-q, n'-3q}
\eeq

Next, we note that the Lagrangian (\ref{glu})
enjoys three discrete interchange
symmetries $u ~:~ \psi_1 ~\leftrightarrow~
\psi_2$; $v ~:~ \psi_2
~\leftrightarrow~ \psi_3$; and $w ~:~ \psi_1
~\leftrightarrow~ \psi_3$; with the
corresponding operations on the two gauge
fields $a$ and $b$. Indeed,
mathematically, the three operations $u$, $v$,
and $w$ generate the
permutation group $S_3$ on three objects, and
our construction amounts to finding
the two-dimensional representation induced on
$a$ and $b$. Taking out the
irrelevant factors we can represent $u$, $v$,
and $w$ on $(x,x')$ (and
similarly on $(y,y')$ as follows. Define the two
dimensional column vector $X$
with the components $x$ and $x'$. Then under
the three discrete interchange symmetries $X
\stackrel{u}{\longrightarrow}
\pmatrix{-1&0\cr 0&1\cr} X$; $X
\stackrel{v}{\longrightarrow} {1 \over
2}\pmatrix{1&3\cr 1&-1\cr} X$; and $X
\stackrel{w}{\longrightarrow}{1 \over
2}\pmatrix{1&-3\cr-1&-1\cr} X$. Imposing
these transformations on the wave
function (\ref{xb}) we obtain the conditions
\beq\label{c1}
c_{nn'} ~=~ c_{-n,n'}
\eeq
\beq\label{c2}
c_{nn'}~=~ c_{{n+n' \over 2}, {3n-n'\over 2}}
\eeq
and
\beq\label{c3}
c_{nn'}~=~ c_{{n-n' \over 2}, -{3n+n'\over 2}}
\eeq
The conditions (\ref{cst}) and (\ref{c1}) imply
that we can restrict $n$ to range
over $(0,1,\ldots,q-1, q)$ and $n'$ to range
over $(1,2,\ldots, 6q)$. It is
convenient then to visualize a $q+1$ by $6q$
toroidal lattice (i.e. one with
periodic boundary conditions) with sites
labelled by $(n,n')$ and on which a particle hops
according to the rules
\beq\label{shop}
(n,n') \stackrel{u}{\longrightarrow} (n-q, n'-3q)
\eeq
\beq\label{vhop}
(n,n') \stackrel{v}{\longrightarrow} {1 \over
2}(n+n', 3n-n')
\eeq
\beq\label{whop}
(n,n') \stackrel{w}{\longrightarrow} {1 \over
2}(n-n', -3n-n')
\eeq
We see from the rules (\ref{vhop}) and
(\ref{whop}) that only even lattice sites
($n \pm n' ~=~ $even) are visited. Starting
from a given site, all the sites
visited by the particle by following an arbitrary
sequence of $u$, $v$, and $w$
hops, before returning to the starting site, are
equivalent. We call the set of
points thus visited a trip.

The desired ground state degeneracy $D$ is
equal to the number of different trips the
particle can take
(or equivalently, the number of inequivalent
sites on the lattice.)

The reader can easily compute $D$ pictorially
for small values of $q$ by drawing a
$(q+1)$ by $6q$ lattice and hop around on it
according to the hopping rules given
above. For example, for $q=3$, we have the
trips \\
$(0,2) ~\sim~(1,17) ~\sim~(2,8)
~\sim~(3,11)$,\\
$(0,4) ~\sim~(2,16)~\sim~(3,13)
~\sim~(1,7)$, \\
$(0,6) ~\sim~(3,15)$,\\
$(0,8) ~\sim~(2,14) ~\sim~(3,17)
~\sim~(1,5)$, \\
$(0,10) ~\sim~(1,13) ~\sim~(2,4)
~\sim~(3,1)$, \\
$(0,12) ~\sim~(3,3)$,\\
$(0,14) ~\sim~(1,11) ~\sim~(2,2)
~\sim~(3,5)$, \\
$(0,16) ~\sim~(2,10) ~\sim~(1,1)
~\sim~(3,7)$, \\
$(0,18) ~\sim~(3,9)$, \\
$(1,3) ~\sim~ (1,15)~\sim~(2,12) ~\sim~(1,9)
~\sim~(2,6) ~\sim~(2,18)$, \\
There are ten trips, one with six sites visited,
six with four sites visited, and three with two
sites visited.  This inventory of the number of
trips with given lengths is also characteristic
of the QH state being studied. Thus,
the topological degeneracy of this QH state is
$D(q=3)~=~ 10$. We can also check that the
total number of sites visited ($=1 \cdot 6+6 \cdot 4+3 \cdot
2= 36)$ is indeed equal to the number of even
lattice sites ${1\over 2} 6 q(q+1) |_{q=3}=36.$

We can readily determine $D$ for an arbitrary
$q$. We argue that we start with a
lattice whose number of sites is a quadratic
function of $q$ and that this number is
reduced by various symmetry relations. So it is
at least plausible that $D(q)$ is
a quadratic of the form $aq^2 + bq +c$. It is
simple to determine $D(q=1) ~=~3$
(Laughlin's result \cite{laugh} !) and
$D(q=2)=6$, in addition to the result we showed
explicitly $D(q=3)=10$. Fitting to these three
points
we find
\beq\label{dd}
D ~=~ {1 \over 2}~(q+1)(q+2)
\eeq
We have verified this result by hand (for $q=4$)
and by a computer program (for a large number of
$q$'s).

We now follow the alternative of gluing the partons into electrons using an
$SU(3)$ gauge field. The effective parton theory is given by
\beq\label{55}
\cl~=~ i \psi^{\dag} (\pa_t -i a_0) \psi + {1
\over 2m}
\psi^{\dag}(\pa_i-i \frac{e}{3} A_i-i a_i)^2 \psi
\eeq
where $\psi=\pmatrix{\psi_1\cr \psi_2\cr
\psi_3\cr}$ and $a_\mu$ are
hermitian traceless 3 by 3 matrices.
Note that (\ref{susan}) is just the restriction of (\ref{55}) to the diagonal
subgroup of $SU(3)$. After integrating out the partons (with each species
filling out $q$ Landau levels), we get
the $SU(3)_q$ CS theory:
\beq\label{SU3CS}
\cl = \frac{ q }{3\times 4\pi} A_\mu\pa_\nu
A_\la \eps^{\mu\nu\la}+
\frac{q}{4\pi}{\rm Tr}\eps^{\mu\nu\la}
(a_\mu \pa_{\nu} a_\la + \frac{i}{3}
a_\mu a_\nu a_\la)
\eeq
Again we choose the $a_0=0$ gauge.
Following what we did for the $SU(2)$ case,
we find the gauge invariant Wilson loops are given in terms of the following
spatially constant gauge potentials (in analogy with (\ref{a123})):
\begin{equation}
\label{a38}
 a_1(x_1,x_2,t)=2\pi \frac{u_1(t)}{L_1} \Lambda_3
+2\pi \frac{u_2(t)}{\sqrt{3}L_1}
\Lambda_8,\ \ \
 a_2(x_1,x_2,t)=2\pi \frac{v_1(t)}{L_1} \Lambda_3
+2\pi \frac{v_2(t)}{\sqrt{3}L_1}
\Lambda_8,\ \ \
\end{equation}
where
\begin{equation}
\Lambda_3 = \pmatrix{1&0&0\cr 0&-1&0\cr 0&0&0&\cr},\ \ \
\Lambda_8 = \pmatrix{1&0&0\cr 0&1&0\cr 0&0&-2&\cr}
\end{equation}
We see that at low energies the non-trivial physical degrees of freedom
of the $SU(3)$ CS theory are described by
$\v u=(u_1,u_2)$ and $\v v=(v_1,v_2)$ where we introduced a
vector notation. After writing the effective Lagrangian for $\v u$ and $\v v$,
we see that
$\v u$  and  $\v v$ satisfy the following commutation relation
\begin{equation}
\label{viuj}
[u_i,v_j]= {i  \delta_{ij} \over 2\pi (2q)}, \ \ \
[u_i, u_j]=0, \ \ \ [v_i,v_j]=0
\end{equation}

The large gauge transformations
\begin{eqnarray}
&& \exp \left(i \frac{2\pi x_1}{L_1} \pmatrix{1&&\cr &-1&\cr &&0\cr} \right)
\nonumber \\
&& \exp \left(i \frac{2\pi x_1}{L_1} \pmatrix{0&&\cr &1&\cr &&-1\cr} \right)
\nonumber \\
&& \exp \left(i \frac{2\pi x_1}{L_1} \pmatrix{-1&&\cr &0&\cr &&1\cr} \right)
\end{eqnarray}
and the ones in the $x_2$ direction lead to the following
equivalence relations:
\begin{equation}
\v u \sim \v u + \v e_i,\ \ \
\v v \sim \v v + \v e_i
\end{equation}
where
$\v e_1 =( 1,0)$,
$\v e_2 =( -\frac{1}{2} , \frac{\sqrt{3}}{2})$, and
$\v e_3 =( -\frac{1}{2} ,-\frac{\sqrt{3}}{2})$.
Note that the angles between the $\v e_i$'s are 120$^\circ$.

{}From the requirement $\psi(\vec{u}) ~=~ \psi(\vec{u}+ \vec{e}_i)$,
$i=1,2,3$, we have
the wave function

\beq\label{w1}
\psi(\vec{u}) ~=~ \sum_{\vec{w}} ~ c_{\vec{w}} ~e^{i 2 \pi \vec{w} \cdot
\vec{u}}
\eeq
where $\vec{w}$ is restricted by demanding that $\vec{w} \cdot \vec{e_i}$ are
integers. From the commutation relation between $\vec{u}$ and its conjugate
momentum
$\vec{v}$, we obtain the wave function

\beq\label{w2}
\phi(\vec{v}) ~=~ \sum_{\vec{w}} ~ c_{\vec{w}}~\delta \left(\vec{v} ~-~ {1
\over 2q}
~\vec{w}\right)
\eeq
Thus, $\phi(\vec{v})$ is non-vanishing only when

\beq\label{w3}
\vec{v} ~=~ {1 \over 2q} ~\vec{w}
\eeq
or upon dotting with $\vec{e_i}$, only when $2q \v v \cdot \v e_i =
\hbox{integers}|_{i=1,2,3}$

Thus the wave function $\phi(\v v)$ is non-zero only on the dual lattice points
spanned by
$(\v d_1,\v d_2)$ which satisfy $\v d_i\cdot \v e_j=\delta_{ij}/2q$.
Due to the periodic condition $\phi(\v v)=\phi(\v v +\v e_i)$
only the circles inside the
``unit cell'' spanned by $\v e_1$ and $\v e_2$ can be independent.
(See Fig. 2)
\begin{figure}[htbp]
\epsfxsize=3in
\begin{center}
\leavevmode
\epsfbox{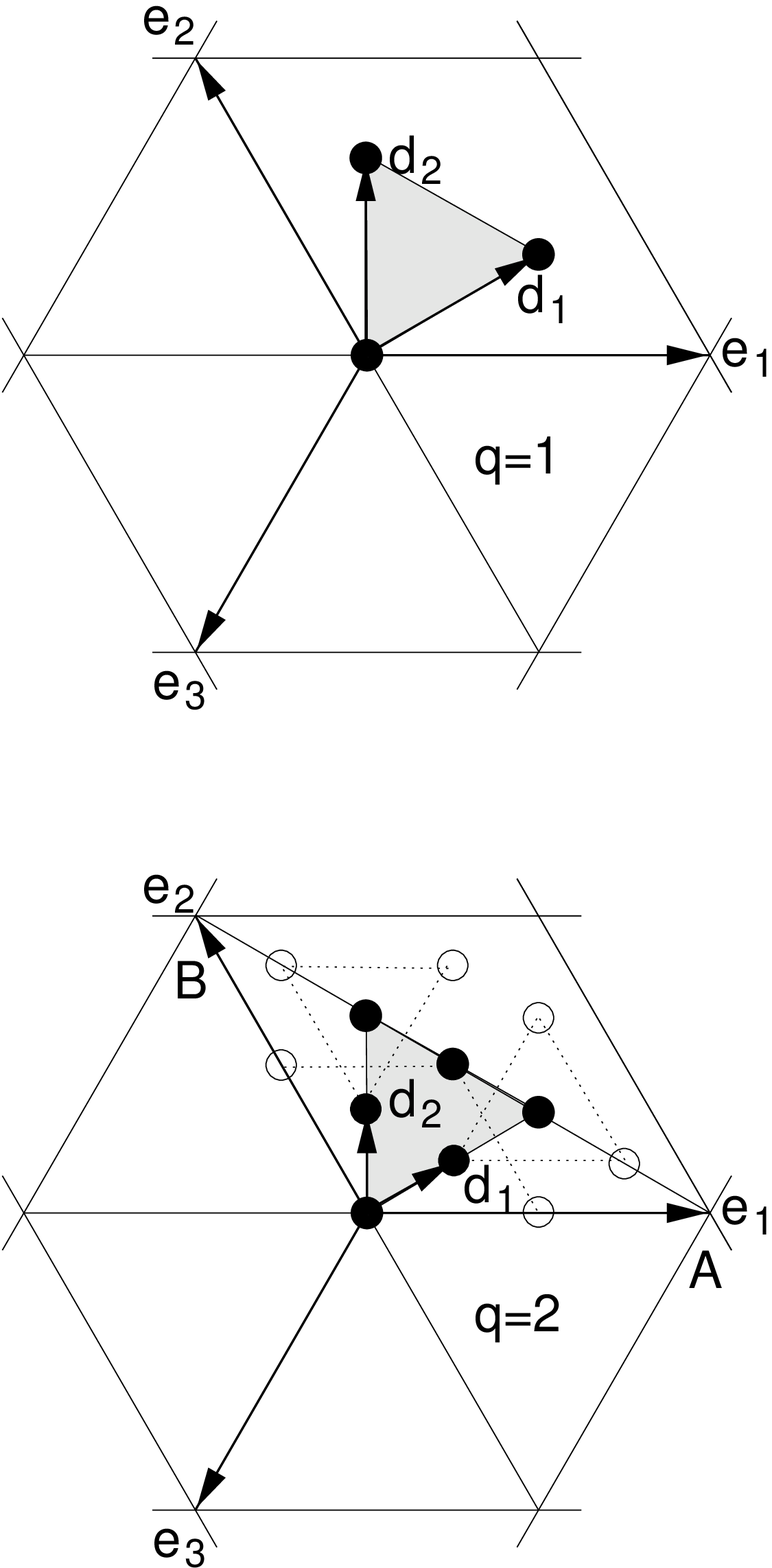}
\end{center}
\caption{
The ground states of $\chi_q^3$ non-abelian state for (a) $q=1$
and (b) $q=2$.}
\label{fig2}
\end{figure}

The global $SU(3)$ gauge transformations
\begin{equation}
\pmatrix{1&&\cr &-1&\cr &&0\cr} \to
\pmatrix{0&&\cr &1&\cr &&-1\cr},\ \ \ \
\pmatrix{0&&\cr &1&\cr &&-1\cr} \to
\pmatrix{-1&&\cr &0&\cr &&1\cr}
\end{equation}
generate a simultaneous 120$^\circ$ rotation on $\v u$ and $\v v$.
This leads to the following equivalence relation:
\begin{equation}
(\v u, \v v) \sim (R_{120^\circ} \v u ,R_{120^\circ} \v v)
\end{equation}
The global $SU(3)$ gauge transformation $(\psi_1,\psi_2,\psi_3)
\to (i\psi_2, i\psi_1,\psi_3)$ also generate a transformation
$(u_1, u_2, v_1, v_2) \to (-u_1, u_2, -v_1, v_2)$ and the corresponding
equivalence relation.

All the above equivalence relations can be satisfied by requiring the wave
function to satisfy
\begin{equation}
\psi(\v v) = \psi(R_{120^\circ} \v v),\ \ \
\psi(v_1,v_2) = \psi(-v_1, v_2)
\end{equation}
Now only the black circles in the
``unit cell'' are independent (see Fig. 2).
The number of the ground states (the black circles) can be calculated
as follows. First the number of the  black circles inside the shaded
triangle (see Fig. 2) is $\frac{q^2}{2} $. However each of the $3(q-1)$
black circles on the edge is only counted as 1/2 in the above calculation.
To include the other half we need to add $3\frac{q-1}{2}$. The 3 black circles
on the corners is only counted as 1/6 each. So we also need to include
a term $3 \frac{5}{6}$. Thus the total number of states is
$\frac{q^2}{2} + 3\frac{q-1}{2} + 3 \frac{5}{6} = \frac{(q+1)(q+2)}{2}$ in
agreement with (\ref{dd}).

\section{A general picture}

{}A general picture emerges from these calculations.
We can now propose the following complete classification scheme
for the ground state structures for a large class of QH states on a torus,
abelian or non-abelian:  For all
QH states reachable by the
parton construction, the ground states on the torus are characterized by a
lattice and the hopping rules
on this lattice.

To be more precise, the ground states are labeled by lattice points
in a vector space. The equivalent lattice points label the same
physical ground state. There are two kinds of equivalence relations:
A) Translation:
\beq
\label{eqA}
\v v \sim \v v + \v e_i |_{i=1,...,{\rm Dim}(\v v)}
\eeq
and B) linear map:
\beq
\label{eqB}
\v v \sim M_i \v v |_{i=1,2,...}
\eeq
where $M_i$ is a ${\rm Dim}(\v v)$ by ${\rm Dim}(\v v)$ matrix.

$\v v$ have conjugate variables, $\v u$, which satisfy the
same equivalence relations (\ref{eqA}) and (\ref{eqB}).
The commutator between $\v v$ and $\v u$ has the form
\beq
\label{uvdel}
[v_i, u_j]= ig_{ij} /2\pi
\eeq
The symmetric matrix $g=(g_{ij})$  defines an inner product
$\v v_1 \cdot \v v_2 \equiv v_{1_i} (g^{-1})_{ij} v_{2_j}$ (which may not be
positive definite).
The lattice that labels the ground states is just
the dual lattice (which will be called the $\v d$-lattice)
of the lattice generated by basis vectors
${\v e_i}$ (which will be called the $\v e$-lattice). 
The basis vectors ${\v d_i}$ of the $\v d$-lattice are given by
\beq
\v d_i \cdot \v e_j = \delta_{ij}
\eeq
For this to be consistent with the equivalence relation (\ref{eqA}),
we also require that the $\v e$-lattice is a sub-lattice of
the $\v d$-lattice. In other words,
$K_{ij}= \v e_i \cdot \v e_j$ must be an integer for all $i$ and $j$.
This also implies that
\beq
\label{didj}
 \v d_i \cdot \v d_j = (K^{-1})_{ij} ,
\eeq
where $K$ is matrix: $K=(K_{ij})$.
We would like to remark that different choices of basis for $\v v$ and $\v u$
lead to different $g_{ij}$. But $K_{ij}$ does not depend on the
choice of the basis. In our discussion of the $SU(3)$ QH liquid, $g_{ij}$
was taken to be $\delta_{ij}/2q$ (see (\ref{viuj})).

For abelian states, only the  equivalence relation (\ref{eqA}) appears and the
$K$-matrix completely describes the ground state structures.
The ground states are labeled by the points on the $\v d$-lattice
inside the unit cell of the $\v e$-lattice.
However, for non-abelian states, we also have an additional type of
equivalence relation of the form in (\ref{eqB}). This completely changes the
structure of the ground states. Only part of the points in the unit cell
correspond to independent ground states.

We see that the ground state structure of a QH state on a torus
is described by a lattice characterized by the $K$-matrix (\ref{didj})
plus a set of linear maps within the lattice.

Now let us discuss what kind of linear maps are allowed.
For convenience, we choose the basis for $\v v$ and $\v u$ such that
$g_{ij}=(K^{-1})_{ij}$. This is always possible since $g_{ij}$ and $K_{ij}$
have the same signature ({\it ie} the same number of positive
eigenvalues and the same number of negative eigenvalues).
With this choice of basis, $\v e_i$ become the standard basis vector: the
$j^{th}$ elements of $\v e_i$ is just $(\v e_i)_j=\delta_{ij}$.

To obtain the condition on the maps $M_i$, first we note that the map $M_i$
acts on both $\v v$ and $\v u$ and keep the commutator (\ref{uvdel})
unchanged. Thus $M_i$ must leave $g_{ij}$ or in our case $K_{ij}$
invariant:
\beq
M_i^T K M_i=K
\label{MKM}
\eeq
This implies that det$(M_i)=\pm 1$. $M_i$ should also map the $\v e$-lattice
onto itself. This requires $M_i$ to be an integer matrix.
Thus the allowed maps $M_i$ are elements in $L({\rm Dim}(K), Z)$
which leaves $K$ invariant (\ref{MKM}).
To obtain the $M_i$'s, we may start with  a transformation between the
partons which leave the electron operator unchanged,
as  we did for the $U(1)_{2q+4} ~\times~ SU(2)_q/Z_2$ state.
Such a transformation induces a transformation on
the gauge fields, and hence a transformation on $\v v$'s and $\v u$'s
 which is nothing but the $M_i$ transformation.

It is interesting to work out the $K$ matrix for the
$U(1)_{2q+4} ~\times~ SU(2)_q/Z_2$ and the  $SU(3)_q$ states.
For the $U(1)_{2q+4} ~\times~ SU(2)_q/Z_2$ state we have
$g=\pmatrix{1/(2q+4) & 0\cr 0& 1/2q\cr}$,
$\v e_1= (1, 0)$
and  $\v e_2= (1/2, -1/2)$ (see (\ref{SU1SU2}), (\ref{e1U1SU2}), and
(\ref{e4U1SU2}) ).  Thus $K=\pmatrix{2q+4 & q+2 \cr q+2 & 1\cr}$.
For the  $SU(3)_q$ state, we have $g=\pmatrix{1/2q & 0\cr 0& 1/2q\cr}$,
$\v e_1= (1, 0)$
and  $\v e_2= (-1/2, \sqrt{3}/2)$.
Thus $K=\pmatrix{2q & -q \cr -q & 2q\cr}$.

In our previous work \cite{rev} we have
emphasized that, for abelian states, the filling
fraction $\nu$, even when supplemented by the
topological degeneracy $D$, cannot (evidently)
capture all
the information contained in the matrix $K$.
Similarly for the non-abelian
states. For example, we note that the
$U(1)_{2q+4} ~\times~ SU(2)_q/Z_2$ state studied in section 4
has degeneracy $D~=~ {1 \over 2} (q+1)(q+2)$,
exactly the same as the degeneracy in (\ref{dd}) for the $SU(3)_q$ states.
However the two states are in general different. For one thing the
$U(1)_{2q+4} ~\times~ SU(2)_q/Z_2$ state has a
filling fraction $\nu ~=~ {q \over q+2}$, while the $SU(3)_q$ state has
$\nu=\frac{q}{3}$. But, for
$q~=~ 1$, we have the same $\nu$ and
$D$, and indeed they both correspond to the
Laughlin $\nu ~=~ {1\over 3}$ state even
though their effective theories are quite
different: $SU(3)_1$ has eight gauge
potentials, while $U(1)_6 \times SU(2)_1$ has
only four.

We trust that the reader can now work out the
lattice and the hopping rules for
any non-abelian states reachable by the parton
construction. For a state
described by the group $G$, the dimensions of
the lattice is given by the rank of
$G$. For example, if we cut the electron into
five equal pieces and construct the
$SU(5)_q$ states, we would have a particle
hopping on a four dimensional lattice.

\section{Summary}

In this paper we propose a simple method to calculate the ground state
degeneracy
of several non-abelian QH liquids. Our method can be applied to any
abelian and non-abelian QH liquids obtained from the parton construction.
A general pattern emerge from our calculation. For the QH liquids reachable by
the parton construction, the ground states on a torus can be described by
points on a lattice.
For abelian QH liquids, the ground states correspond to points inside a
``unit cell''.
For non-abelian states the ground states correspond to points inside
a ``folded'' unit cell. The folding is generated by reflection, and
possibly rotations. (See Fig. 1 and Fig. 2)

\section{Acknowledgements}

XGW is supported by NSF Grant No. DMR--97--14198 and AZ by NSF Grant No.
PHY89-04035. AZ thanks L. Balents for help with
computer programming.

\section{Appendix: Coset Construction}

In this Appendix, we are going to study the coset construction
of the CS theory. We will see that the coset construction is closely related
to the parton construction.

We start with a physical example.
The $\nu=1/k$ Laughlin state can be constructed from a parton
construction \cite{Jain}.
For example, to construct the $\nu=1/2$ Laughlin state (of bosons),
we may
start with two kinds of partons in the $\nu=1$ state, with the wave
function $\Psi \sim
\prod_{ij} (z^{(1)}_i- z^{(1)}_j) \prod_{ij} (z^{(2)}_i- z^{(2)}_j)$ where
$z^{(1)}_i$ and $z^{(2)}_i$ represent the coordinates of the two kinds of
partons, described by
a CS effective theory (\ref{cs}) with $K=\pmatrix{1&0\cr 0&1\cr}$.
This CS effective
theory is thus the $U(1)_1\times U(1)_1$ CS theory.

Suppose we now make a projection by setting $z^{(1)}_i= z^{(2)}_i$.
Physically, we
recombine the two kinds of partons together into electrons.
The wave function becomes $\Psi
\sim \prod_{ij} (z_i- z_j)^2$ and the parton state becomes the
$\nu=1/2$ Laughlin state.

We will now describe how this
projection can be achieved in the
$U(1)_1\times U(1)_1$ CS effective theory.
Note that the projection binds the two kinds of partons so that they always
move together.
Thus after projection, their currents and
densities $j^{(1)}_\mu=\frac{1}{2\pi} \eps_{\mu\nu\la} \pa_\nu a^{(1)}_\la$
and $j^{(2)}_\mu=\frac{1}{2\pi} \eps_{\mu\nu\la} \pa_\nu a^{(2)}_\la$
must be equal. This leads to the constraint $ a^{(1)}_\mu - a^{(2)}_\mu=0$.
By setting $a^{(1)}_\mu= a^{(2)}_\mu$ in $\cl =~ {1 \over 4\pi}
(~\eps^{\mu \nu \la} ~ a_\mu^{(1)}
~\partial_\nu ~a_\la^{(1)}+ ~\eps^{\mu \nu \la} ~ a_\mu^{(2)}
~\partial_\nu ~a_\la^{(2)})$ we see that the $U(1)_1\times U(1)_1$ CS
theory is reduced to the
$U(1)_2$ CS theory which is the effective theory of the $\nu=1/2$ Laughlin
state.

Next we would like to understand how the parton-construction
realizes itself in the ground
states of the QH liquids. In other words,  we would like to start with the
ground
states of the  $U(1)_1\times U(1)_1$ CS theory,
and to see how to
``project'' to obtain the ground states of the $\nu=1/2$ Laughlin state
({\it i.e.} the  ground states of the $U(1)_2$ CS theory). It turns out that
this can be achieved through the coset construction.

We know that before projection, the edge excitations of the parton
state (described by the  $K=\pmatrix{1&0\cr 0&1\cr}$ CS theory) is described
by two currents $j_{1,2}$ associated with each kind of partons. The currents
form two level-1 $U(1)$ Kac-Moody (KM) algebras: $U(1)_1\times U(1)_1$. The
relative motion of the two kinds of partons is described by $j_-=j_1-j_2$
which forms a level-2 $U(1)$ KM algebra: $U(1)_2$. (Schematically,
$[j_1,j_1]\sim 1$ and $[j_2,j_2]\sim 1$ and hence
$[j_1-j_2,j_1-j_2]\sim 2$).
After projecting out the
the $U(1)_2$ KM algebra, the resulting edge theory is described by
the coset model $U(1)_1\times U(1)_1/U(1)_2$ with a $U(1)_2$ KM algebra
generated by $j_+=j_1+j_2$.

Motivated by the relation between the edge theory
and the CS theory, we would like to study the coset theory of the CS theory.
In particular, we would like to show that the coset CS theory
$U(1)_1\times U(1)_1/U(1)_2$ is nothing but the $U(1)_2$ CS theory, as
suggested by the parton construction given above.

Let us first start with a simpler problem. Consider a $U(1)_k$ CS theory,
which has $k$ degenerate ground states on the torus. The coset $U(1)_k/U(1)_k$
model can be obtained by making
the projection $a_\mu=0$. Obviously, after projection there is no non-trivial
low energy excitation and the ground state is non-degenerate.

The ground states of the $U(1)_k$ CS theory are described by
$x$ and $y$ operators
satisfying the commutator (\ref{one*}). First we note that $x$ and $y$
operators
do not satisfy the equivalence condition (\ref{equiv}), and hence are not
operators that act within the the physical Hilbert space. The allowed
operators are $ U_{mn}\equiv e^{2 \pi i(m x+n y)}$ with $m,n=$integers.
Naively one may want to project by requiring the states to satisfy
$x \psi=y \psi=0$. Or more precisely, one must require
$U_{10}\psi=U_{01}\psi=\psi$.
Since $x$ and $y$ (or
$U_{10}$ and $U_{01}$) do not commute (with a commutator equal to a
number), no state satisfies this
condition. Thus no state survives the projection. This contradicts our
physical intuition that there should be one and only one state that
survives the
projection.

To construct the projected theory $U(1)_k / U(1)_k$ correctly,
we start with a $U(1)_k  \times U(1)_{-k}$ theory described by the
effective Lagrangian
\beq\label{dou}
L ~=~ 2 \pi ~ k ~(\dot{x} ~y ~-~ \dot{x}' ~y')
\eeq
describing two particles living on the torus. The $U(1)_k$ in $U(1)_k ~\times~
U(1)_{-k}$ is the theory
before projection and $U(1)_{-k}$ is the ``conjugate'' of the projection
and hence a $U(1)_{-k}$ theory. (In general, to construct the $G/H$ coset
model, one may start with the $G\times H^*$ model.) We have
\beq\label{x}
\left[x,y \right] ~=~ {i \over 2 \pi k}
\eeq
and
\beq\label{xp}
\left[x',y' \right] ~=~ -{i \over 2 \pi k}
\eeq
thus implying
\beq\label{xxp}
\left[ x-x', ~y-y' \right] ~=~ 0
\eeq
As in (\ref{two}) and (\ref{john}), the wave functions are given by
\beq\label{py}
\psi(y,y') ~=~ \sum\limits_{n,n'} ~c_{nn'} ~ e^{i ~2 \pi(ny ~+~ n'y')}
\eeq
and
\beq\label{px}
\phi(x,x') ~=~ \sum\limits_{n,n'} ~c_{nn'} ~ \delta(kx~-~n) ~\delta(-kx'
~-~ n')
\eeq
Periodicity $x\sim x+1$ and $x'  \sim x' ~+~ 1$ imply, as in (\ref{nk}),
\beq\label{per}
c_{nn'} ~=~ c_{n+k, ~ n'} ~=~ c_{n, n' ~+~ k}
\eeq
This leads to a total of $k\times k$ ground states before the projection.

We now project. Remember that we want to impose $x=y=0$. But this is impossible
since $x$ and $y$ do not commute. With the help of the additional sector
$U(1)_{-k}$, we can impose the conditions
\beq\label{blind}
x~-~x' ~=~ 0
\eeq
and
\beq\label{blind2}
y~-~y' ~=~ 0
\eeq
This is possible since $x-x'$ and $y-y'$ commute.
In the language of (\ref{dou}) the projection has the
physical interpretation of binding the two particles together.
Since the coordinates on the torus are defined only mod integer, the right
hand sides of (\ref{blind}) and (\ref{blind2}) should be interpreted as $0$
mod integer. (Or more precisely, we can only impose the condition on
the allowed operators: $e^{i2\pi (x-x')}= e^{i2\pi (y-y')}= 1$.)
Thus the states in the projected theory satisfy
\begin{equation}
 e^{i2\pi (x-x')} \psi= e^{i2\pi (y-y')} \psi = \psi
\end{equation}
Writing (\ref{px}) as
\beq\label{px2}
\psi(x,x') ~=~ \sum\limits_{n,n'} ~c_{n,n'} ~ \delta \left(k(x-x')
~-~(n+n') \right) ~ \delta (kx' ~+~ n')
\eeq
we see that the $c_{nn'}$'s are non-zero only when
\beq\label{md}
n~+~n' ~=~ 0 ~mod~k .
\eeq
Next, writing (\ref{py}) as
\beq\label{nextwriting}
\psi(y,y') ~=~ \sum\limits_{n,n'} ~c_{n,n'} ~ e^{i~2 \pi \left((n~+~n')
~y~+~n'~(y'~-~y) \right)}
\eeq
we see, referring to (\ref{del}), that in order for $\psi(y,y')$ to be
proportional to $\delta^P(y-y')$ as required by (\ref{blind2}), the
$c_{nn'}$'s should depend only on $n~+~n'$:
\beq\label{cd}
c_{nn'}~=~ d_{n+n'}
\eeq
Referring to (\ref{md}) and to the periodicity condition (\ref{per}) we see
that $d_{n+n'}$ is independent of $n+n'$. All the non-zero $c_{nn'}$'s are
equal. There is only one state left after projection, as expected for the
$U(1)_k/U(1)_k$ coset model. Thus, in contrast to the naive construction
described before, we have now managed to obtain the correct result using
this construction
starting with the $U(1)_k ~\times~
U(1)_{-k}$ theory.

Using the fact that $n+n' ~=~ jk$ has to be a multiple of $k$, we can now
obtain
\beqra\label{pyy}
\psi(y,y') &= &\sum\limits_j ~e^{i~2 \pi ~j~k~y} ~ \sum\limits_{n'} ~e^{i~2
\pi n'~(y-y')} \nonumber \\
&= &\delta^P ~(ky) ~ \delta^P ~(y-y')
\eeqra
Similarly, we can write (\ref{px2}) as
\beqra\label{pxx2}
\phi(x,x') &=  &\sum\limits_j ~ \delta \left( k(x-x') ~-~ kj \right) ~
\sum\limits_{n'} ~ \delta(kx' ~+~ n') \nonumber \\
&= &\delta^P (kx') ~ \delta^P (x-x')
\eeqra
We see explicitly that $\phi(x,x')$ and $\psi(y,y')$ have the same status.
Indeed, note
\beq\label{ide}
\delta^P (kx) ~ \delta^P(x-x') ~=~ \delta^P(kx') ~\delta^P(x-x')
\eeq
We remark in passing the obvious fact that $\delta^P(kx) \delta^P(x-x')$
is not equal to $\delta^P(kx) ~\delta^P(kx')$

The above example suggests that to construct a $G/H$ coset model, one may
start with an enlarged theory $G\times H^*$ and then project out a diagonal
part $H\times H^*$ to end up with $G/H$. Such an approach at least works
for the coset construction of KM algebras.

Now we are ready to consider the $U(1)_1\times U(1)_1/U(1)_2$ coset model.
Let us study the more general $U(1)_k\times U(1)_{k'}/U(1)_{k+k'}$ coset
model where $k$ and $k'$ have no common factors.
Starting from the
$U(1)_k\times U(1)_{k'}\times U(1)_{-(k+k')}$ CS theory described by the
operators
$(x,y)$, $(x',y')$, and $(x'',y'')$:
\beq
 [x,y]={i\over 2\pi k},\ \ \ [x',y']={i\over 2\pi k'},\ \ \
[x'',y'']=-{i\over  2 \pi (k+k')},
\eeq
we project by imposing the conditions
\beq\label{blind3}
kx~+~k' x' ~+ ~ k''x'' =~ 0
\eeq
and
\beq\label{blind23}
ky~+~k' y' ~+ ~ k'' y'' =~ 0
\eeq
where we have defined for notational convenience $k''\equiv-(k+k')$.
This is possible since $kx~+~k' x' ~+ ~ k''x''$ and $ky~+~k' y' ~+ ~ k'' y''$
commute.
Again the right
hand sides of (\ref{blind3}) and (\ref{blind23}) should be interpreted as $0$
mod integer, because the coordinates $(x,y)$ etc. are only defined up to an
integer.

Extending (\ref{py}) and (\ref{px}), we write the wave functions as
\beq\label{py1}
\psi(y,y',y'') ~=~ \sum\limits_{n,n',n'} ~c_{nn'n''} ~ e^{i ~2 \pi(ny ~+~
n'y'~+~n''y'')}
\eeq
and
\beq\label{px1}
\phi(x,x',x'') ~=~ \sum\limits_{n,n',n''} ~c_{nn'n''} ~ \delta(kx~-~n)
~\delta(k'x'
~-~ n') ~\delta(k''x''
~-~ n'')
\eeq
Periodicity $x\sim x+1$ etc. imply, as in (\ref{per}),
\beq\label{per1}
c_{nn'n''} ~=~ c_{n+k, ~ n'n''} ~=~ c_{n, n' ~+~ k,n''}~=~ c_{n, n',n''+k''}
\eeq
This leads to a total of $kk'k''$ states before projection.

It turns out that the projection does not remove any states.
Indeed, we see from (\ref{px1}) that (\ref{blind3}) is already satisfied,
posing no restriction on the $c_{nn'n''}$'s. Defining $m=[n]_{k}$ as before
and the corresponding primed and double primed quantities $m'$ and $m''$
we write (Cf (\ref{four})
\beq\label{four2}
\psi(y,y',y'')~ =~ \left(\sum_{m,m',m''=1}^{k,k',k''} ~c_{m,m',m''} ~
e^{i 2\pi (my+m'y'+m''y'')} \right)~ \delta^P(ky) \delta^P(k'y')
\delta^P(k''y'')
\eeq
We find that (\ref{blind23}) does not impose any further restriction either.
Thus the $U(1)_k\times U(1)_{k'}/U(1)_{k+k'}$ coset model has $kk'(k+k'')$
degenerate ground states. In particular, the
$U(1)_1\times U(1)_k/U(1)_{k+1}$ coset model has $k(k+1)$ degenerate ground
states, in agreement with a previous result obtained using a more abstract
and algebraic
approach\cite{coset}.  When $k=1$ the $U(1)_1\times U(1)_1/U(1)_{2}$ coset
model
has $2$ degenerate
ground states which is the same as the number of ground states in the $U(1)_2$
theory and the $\nu=1/2$ Laughlin state.

\pagebreak

\end{document}